\documentclass[twocolumn,showpacs,amssymb,aps,prd,floatfix]{revtex4}
\usepackage{amsmath}
\usepackage{amssymb}
\usepackage{graphicx}
\newcommand{\be}{\begin{equation}}
\newcommand{\ee}{\end{equation}}

\begin{document}

\title{Toroidal marginally outer trapped surfaces in closed Friedmann--Lema\^{\i}tre--Robertson--Walker spacetimes: Stability and isoperimetric inequalities}

\author{Patryk Mach}\email{patryk.mach@uj.edu.pl}
\affiliation{Institute of Physics, Jagiellonian University, \L ojasiewicza 11,  30-438 Krak\'ow, Poland}
\author{Naqing Xie}\email{nqxie@fudan.edu.cn}
\affiliation{School of Mathematical Sciences, Fudan University, Shanghai 200433, China}


\begin{abstract}
We investigate toroidal Marginally Outer Trapped Surfaces (MOTS) and Marginally Outer Trapped Tubes (MOTT) in closed Friedmann--Lema\^{\i}tre--Robertson--Walker (FLRW) geometries. They are constructed by embedding Constant Mean Curvature (CMC) Clifford tori in a FLRW spacetime. This construction is used to assess the quality of certain isoperimetric inequalities, recently proved in axial symmetry. Similarly to spherically symmetric MOTS existing in FLRW spacetimes, the toroidal ones are also unstable.
\end{abstract}

\pacs{04.20.Cv, 04.20.Gz}
\keywords{}
\maketitle

\section{Introduction}

In the existing literature Marginally Outer Trapped Surfaces (MOTS) appear usually in connection with the apparent horizon --- a key concept in an attempt to construct a quasi-local definition of a black hole. Let the spacetime $\mathcal M$ be foliated by a family of spacelike Cauchy hypersurfaces $\{ \Sigma_t \}$. An apparent horizon is defined as a collection of boundaries of regions which contain trapped surfaces in $\{ \Sigma_t \}$. If an apparent horizon is regular, it is foliated by MOTS.

So far the majority of known analytic examples of MOTS was found in spherical symmetry, but there are a few interesting examples of non-spherical ones \cite{mk, ibd, flores}. In this paper we focus on toroidal MOTS. The existence of black holes which, during early stages of their evolution, could have toroidal topology was suggested already in early 1990's \cite{HKWWST, STW, siino1, siino2, siino3}. Toroidal MOTS enclosed within an apparent horizon of spherical topology were constructed numerically in \cite{Husa96}.

In \cite{flores} Flores, Haesen and Ortega constructed a family of toroidal MOTS by embedding Constant Mean Curvature (CMC) Clifford tori in closed Friedmann--Lema\^{\i}tre--Robertson--Walker (FLRW) geometries. Such surfaces can be found analytically during the entire evolution, forming the so-called Marginally Outer Trapped Tubes (MOTT). Of course, their existence is not connected with black holes, but they provide an excellent testbed for various theorems concerning MOTS.

In \cite{kmmmx} we explicitly constructed examples of toroidal MOTS in the class of time-symmetric initial data --- the so-called ``stars of constant density'' \cite{niall}. They were all contained within a spherical black hole. A ``star of constant density'' consists of a spherical region, isometric to a fragment of a 3-sphere, and an external part representing a slice in an appropriately chosen Schwarzschild spacetime. Toroidal MOTS discussed in \cite{kmmmx} fall naturally into two classes: those embedded entirely in the 3-spherical region occupied by the ``star'', and those laying partially in the ``star'' and partially in the Schwarzschild region. Marginally outer trapped surfaces belonging to the first family are precisely Clifford tori that were discussed in \cite{flores} in the context of FLRW spacetimes.

In this paper we investigate Clifford CMC tori, embedded in FLRW spacetimes, as discussed by \cite{flores}. The construction introduced in \cite{flores} uses the Hopf map; here we follow a more direct approach, basing on the stereographic projection and toroidal coordinates, as introduced in \cite{kmmmx}. We then focus on two issues. Firstly we show that the constructed toroidal MOTS are unstable. This fact was suggested in \cite{flores}. It can be thought of as a natural consequence of the SO(4) symmetry of the standard hypersurfaces of constant time, but a precise statement concerning stability of MOTS requires caution. In this work we adopt definitions of the stability of MOTS introduced in \cite{ams,ams2}. Secondly we test certain isoperimetric inequalities recently introduced in \cite{khuri_xie}, and proved in axial symmetry. The quality of some of them, valid for minimal surfaces, was already assessed in \cite{kmmmx}, where we dealt with time-symmetric initial data. Toroidal MOTS and MOTT in closed FLRW cosmological models provide a possibility for a test in the dynamical setting.

We use the gravitational system of units with $c = G = 1$. The signature of the metric tensor is assumed to be $(-,+,+,+)$. Throughout this paper Greek indices are used to label spacetime dimensions. Latin indices are reserved for objects on sections: we denote 3-dimensional objects with lowercase Latin indices, and 2-dimensional objects with capital ones.

\section{Closed FLRW universe}

The metric of a closed FLRW model can be written in the form
\begin{equation}
\label{flrw1}
g = -dt^2 + S^2(t) \left( d\chi^2 + \sin^2 \chi d\Omega^2 \right),
\end{equation}
where $d\Omega^2 = d\theta^2 + \sin^2 \theta d \varphi^2$ denotes the round metric on a 2-sphere, and the so-called scale factor $S = S(t)$ satisfies Friedmann equations. The surfaces of constant time $\Sigma_t$ are round 3-spheres with $0 \le \chi \le \pi$. They are characterized by a 3-dimensional scalar curvature $^{(3)}R = 6/S^2$, constant within each time-slice. The extrinsic curvature of $\Sigma_t$ reads $K_{ij} = - \frac{\dot S}{S} \gamma_{ij}$, where the dot denotes the derivative with respect to time $t$, and $\gamma$ is the induced metric on $\Sigma_t$,
\[ \gamma = S^2(t) \left( d\chi^2 + \sin^2 \chi d\Omega^2 \right). \]
The trace of the extrinsic curvature is $\mathrm{tr} K = \gamma^{ij}K_{ij}  = - 3 \dot S/S$. The Hamiltonian constraint equation
\[ ^{(3)}R - [K_{ij}K^{ij} - (\mathrm{tr}K)^2 ] = 16 \pi \rho \]
yields the expression for the energy density on a given time-slice as
\[ \rho = \frac{3(1 + \dot S^2)}{8 \pi S^2}. \]

Assuming that the matter consists of dust, one gets the solution for the scale factor in the form
\[ S = \frac{S_m}{2}(1 + \cos \eta), \quad t = \frac{S_m}{2}(\eta + \sin \eta), \]
where $S_m$ is a constant. Note that $d \eta = dt/S$. The so-called conformal time $\eta$ changes from $-\pi$ to $\pi$, and $\eta = 0$ corresponds to a maximum in the scale factor ($S = S_m$). Another textbook solution can be obtained for the radiation-dominated fluid with the pressure $p = \rho/3$. In terms of the conformal time $\eta$, it reads
\[ S = S_m \cos \eta, \quad t = S_m \sin \eta. \]
Here $-\pi/2 \le \eta \le \pi/2$.

In the following, it will be convenient to introduce new coordinates on slices $\Sigma_t$ so that the metric induced on each $\Sigma_t$ can be written in a manifestly conformally flat form. This choice is motivated by our previous analysis presented in \cite{kmmmx}.  The (useful) freedom in choosing new coordinates $(T,R)$ is reduced to
\[ t = T, \quad \chi = \chi(T,R). \]
We retain the same coordinates $(\theta, \varphi)$ in $d\Omega^2$. The requirement that each time-slice should be explicitly conformally flat yields a solution for $\chi$ in the form
\[ \chi(T,R) = 2 \arctan (C_1(T) R), \]
where $C_1$ is an arbitrary function of $T$. The metric $g$ can be now written as
\begin{eqnarray}
g & = & - \left[1 - \frac{4 R^2 S^2 \dot C_1^2}{(1 + R^2 C_1^2)^2} \right] dT^2 + \frac{8 R S^2 C_1 \dot C_1}{(1 + R^2 C_1^2)^2} dT dR \nonumber \\
& & + \frac{4 C_1^2 S^2}{(1 + R^2 C_1^2)^2} (dR^2 + R^2 d\Omega^2),
\label{conformal}
\end{eqnarray}
where the dot denotes the derivative with respect to $T$.

Notice that the transformation $\chi = 2 \arctan (C_1(T) R)$ defines a stereographic projection from the 3-sphere to $\mathbb R^3$. The freedom in choosing the value of $C_1$ is simply equivalent to the rescaling of the stereographic projection. In our setting, the 3-sphere of unit radius is projected from the pole corresponding to $\chi = \pi$ to the equatorial hyperplane, if we choose $C_1 = 1$. For simplicity, we will further assume $C_1 = \mathrm{const}$. This gives the metric in the form
\begin{equation}
\label{g1}
g = - dT^2 + \Phi^4(R) (dR^2 + R^2 d\Omega^2),
\end{equation}
where the spatial conformal factor reads
\begin{equation}
\label{g2}
\Phi(R) = \frac{\sqrt{2 C_1 S}}{\sqrt{1 + R^2 C_1^2}}.
\end{equation}

\section{Toroidal MOTS in the closed FLRW universe}

We will now construct toroidal MOTS in a closed FLRW universe. They belong to the family of the so-called generalized (or CMC) Clifford tori.

Let us choose a hypersurface of constant time $\Sigma_t$. A future pointing unit vector normal to $\Sigma_t$ will be denoted by $n^\mu$. Let $\mathcal S$ be a closed 2-surface in $\Sigma_t$, and let $m^\mu$ denote an outward-pointing unit vector normal to $\mathcal S$ and tangent to $\Sigma_t$. We define two null vectors: $l_\pm^\mu = n^\mu \pm m^\mu$. The two expansion scalars associated with $l_\pm^\mu$ are defined as
\[ \theta_\pm = \pm H - K_{ij}m^i m^j + \mathrm{tr} K, \]
where $H$ denotes the mean curvature of $\mathcal S$,
\[ H = \nabla_i m^i = \frac{1}{\sqrt{\mathrm{det} \gamma}} \partial_i \left( \sqrt{\mathrm{det} \gamma} m^i \right). \]
Here $\nabla_i$ is the covariant derivative associated with the induced metric on $\Sigma_t$, i.e., $\gamma = \Phi^4(R)(dR^2 + R^2 d\Omega^2)$. Note that $K_{ij}m^i m^j = - \frac{\dot S}{S} \, \gamma_{ij} m^i m^j = - \frac{\dot S}{S}$. Accordingly
\[ \theta_\pm = \pm H - \frac{2 \dot S}{S}. \]

A surface $\mathcal S$ is called outer trapped if $\theta_+ < 0$ everywhere on $\mathcal S$. If $\theta_+ = 0$, the surface $\mathcal S$ is called a MOTS. Here the term ``outer'' refers to a particular choice of the direction $l_+^\mu$. A surface $\mathcal S$ for which $H = 0$ is called a minimal one. Note that for time-symmetric data with $K_{ij} = 0$, minimal surfaces coincide with MOTS.

We now work in toroidal coordinates $(\sigma, \tau, \phi)$ within $\Sigma_t$. They are related to the Cartesian coordinates $(x,y,z)$ by
\begin{eqnarray*}
x & = & \frac{c \sinh \tau \cos \phi}{\cosh \tau - \cos \sigma}, \\
y & = & \frac{c \sinh \tau \sin \phi}{\cosh \tau - \cos \sigma}, \\
z & = & \frac{c \sin \sigma}{\cosh \tau - \cos \sigma}.
\end{eqnarray*}
The relation between the toroidal coordinates $(\sigma, \tau, \phi)$ and the spherical coordinates $(R,\theta,\varphi)$ used in this paper is
\begin{eqnarray*}
R & = & \frac{c \sqrt{\sinh^2 \tau + \sin^2 \sigma}}{\cosh \tau - \cos \sigma}, \\
\cot \theta & = & \frac{\sin \sigma}{\sinh \tau}, \\
\varphi & = & \phi.
\end{eqnarray*}
Here $- \pi \le \sigma \le \pi$, $\tau \ge 0$, $0 \le \phi < 2 \pi$, and $c > 0$ is a radius of the circle in the $z = 0$ plane corresponding to $\tau = \infty$. In terms of coordinates $(\sigma, \tau, \phi)$, the flat Euclidean metric can be expressed as
\begin{eqnarray*}
\lefteqn{dx^2 + dy^2 + dz^2 = dR^2 + R^2 d\Omega^2 = }\\
&& \frac{c^2}{(\cosh \tau - \cos \sigma)^2} \left( d \sigma^2 + d \tau^2 + \sinh^2 \tau d \phi^2  \right).
\end{eqnarray*}
Let us choose $c = 1/C_1$. Such a choice yields a particularly simple form of the metric $g$ in coordinates $(T,\sigma,\tau,\phi)$:
\begin{equation}
\label{gtoroidal}
g = -dT^2 + S^2 \mathrm{sech}^2 \tau \left( d\sigma^2 + d\tau^2 + \sinh^2 \tau d\phi^2 \right).
\end{equation}
Note that the spatial part of the metric is still manifestly conformally flat.

Consider a torus $\mathcal S$ defined by setting $\tau = \mathrm{const}$. An outward pointing unit vector, normal to $\mathcal S$ has the components
\[ m^i = (m^\sigma,m^\tau,m^\phi) = (0,-\cosh \tau/S, 0). \]
The mean curvature of $\mathcal S$ reads
\[ H = \frac{(\cosh (2 \tau )-3) \text{csch}(\tau )}{2 S}. \]
This expression does not depend on $\sigma$. Accordingly, each torus of constant $\tau$ happens to be a Constant Mean Curvature (CMC) surface. One gets $H = 0$ (a minimal surface) for $\tau = \tau_0 = \mathrm{arcosh}(3)/2 = \log (1 + \sqrt{2})$, independently of the value of $S$. Consequently, the minimal torus remains fixed (with respect to the 3-sphere $\Sigma_t$) during the entire evolution.

The scalar expansion $\theta_+$ of a torus of constant $\tau$ is now simply
\[ \theta_+ = \frac{(\cosh (2 \tau )-3) \text{csch}(\tau )}{2 S}  - \frac{2 \dot S}{S}. \]
A torus with $\tau$ corresponding to a solution of the condition $\theta_+ = 0$, i.e.,
\begin{equation}
\label{eqFLRW}
(\cosh (2 \tau )-3) \text{csch}(\tau ) - 4 \dot S = 0
\end{equation}
is therefore a MOTS. The only solution of Eq.\ (\ref{eqFLRW}) satisfying $\tau > 0$ is
\begin{equation}
\label{solFLRW}
\tau = \mathrm{arsinh} \left( \dot S + \sqrt{1 + \dot S^2} \right).
\end{equation}
This provides a general description of toroidal MOTS in closed FLRW geometries. Note that for the particular case of the FLRW universe filled with dust
\begin{equation}
\label{dustSd}
\dot S = - \frac{\sin \eta}{1 + \cos \eta}.
\end{equation}
Substituting Eq.\ (\ref{dustSd}) into Eq.\ (\ref{solFLRW}) one obtains, for the dust solution,
\[ \tau = \mathrm{arsinh} \left[ \frac{2 \cos (\eta/2) - \sin \eta}{1 + \cos \eta} \right]. \]
The corresponding expression in the case of the radiation-dominated FLRW universe is even simpler. One gets $\dot S = - \tan \eta$ and
\[ \tau = \mathrm{arsinh} \left( \frac{1 - \sin \eta}{\cos \eta} \right). \]

\begin{figure}
\includegraphics[width=\columnwidth]{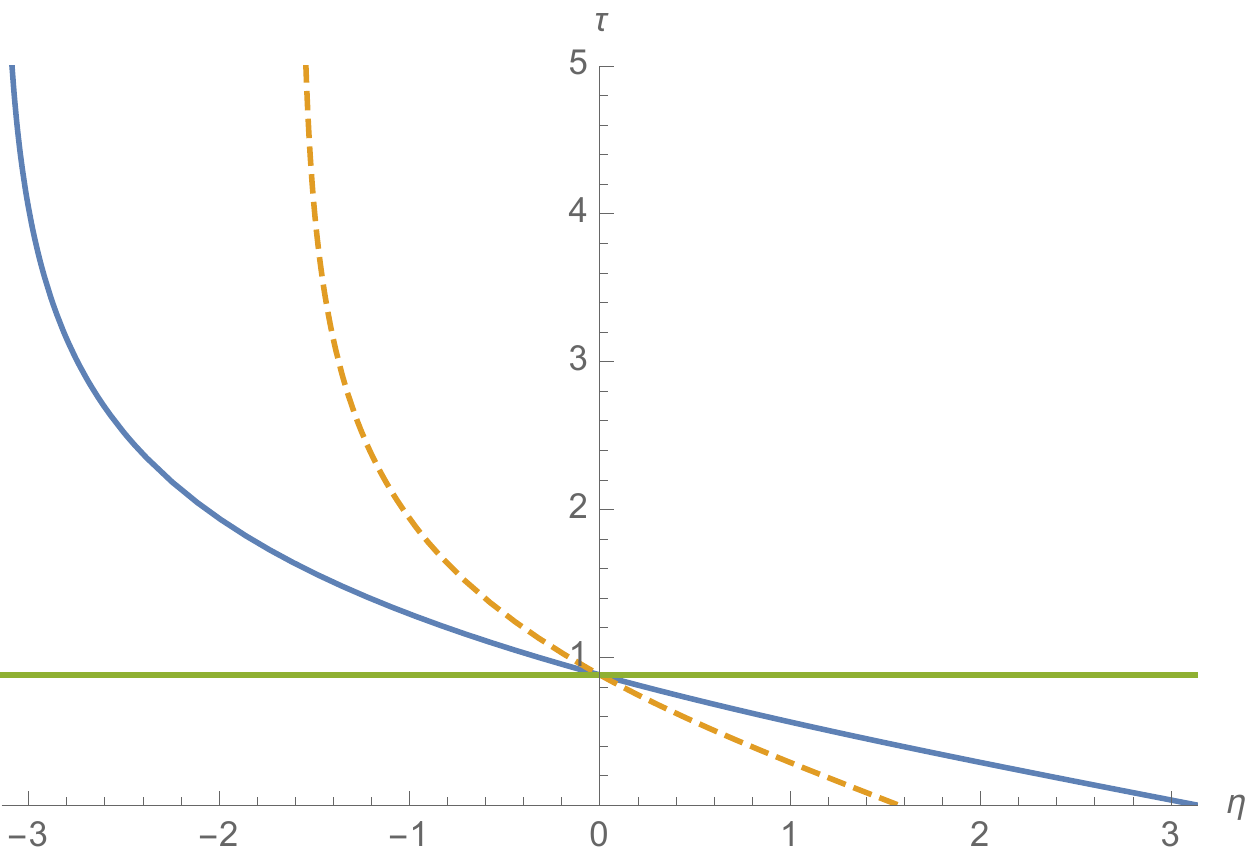}
\caption{\label{fig1}The plot of solution (\ref{solFLRW}) for the dust (solid blue line) and radiation-dominated (dashed line) FLRW models. The horizontal line marks the value of $\tau = \tau_0 = \log(1 + \sqrt{2})$ corresponding to a minimal surface.}
\end{figure}

Figure \ref{fig1} shows the graph of $\tau$ as a function of the conformal time $\eta$ for the dust and radiation-dominated FLRW models. In the dust case the parameter $\tau$ drops from infinity to $\tau = \tau_0 = \log (1 + \sqrt{2})$, as $\eta$ goes from $-\pi$ to 0, and then drops further from $\tau=\tau_0$ to $\tau = 0$, as $\eta$ goes from 0 to $\pi$. This corresponds to an infinitely thin torus for $\eta = -\pi$ which grows during the entire cycle of evolution until $\eta = \pi$. For $\eta = 0$ ($K_{ij} = 0$), the marginally outer trapped torus coincides with the minimal one, as expected. The behavior in the radiation-dominated case is analogous, except for the conformal time changing from $-\pi/2$ to $\pi/2$.

Constant mean curvature tori constructed above are sometimes called the generalized Clifford tori (we will use the term CMC Clifford tori in this paper). The simplest way to define them is to consider the Euclidean space $\mathbb R^4$ with coordinates $(x_1,x_2,x_3,x_4)$, and a surface given by two conditions
\begin{eqnarray}
x_1^2 + x_2^2 & = & r^2, \nonumber \\
x_3^2 + x_4^2 & = & 1 - r^2,
\label{cliffordcmc}
\end{eqnarray}
where $0 < r < 1$ is a constant. Clearly, such a torus is embedded in the unit 3-sphere $\mathbb S^3$. In hyperspherical coordinates $(\chi,\theta,\phi)$ defined as
\begin{eqnarray*}
x_1 & = & \sin \chi \sin \theta \cos \phi, \\
x_2 & = & \sin \chi \sin \theta \sin \phi, \\
x_3 & = & \sin \chi \cos \theta, \\
x_4 & = & \cos \chi,
\end{eqnarray*}
Eqs.\ (\ref{cliffordcmc}) read
\begin{eqnarray*}
\sin^2 \chi \sin^2 \theta & = & r^2, \\
\sin^2 \chi \cos^2 \theta + \cos^2 \chi & = & 1 - r^2.
\end{eqnarray*}
Assuming $c = C_1 = 1$, and performing the stereographic projection given by $\chi = 2 \arctan R$, we see that our CMC tori satisfy the above equations with $r = \tanh \tau$.

The minimal torus with $\tau = \tau_0 = \log (1 + \sqrt{2})$ or $r = 1/\sqrt{2}$ is known as the Clifford torus.

The theory of closed minimal surfaces embedded in the 3-sphere is a classic, but still active field in differential geometry. In 1970 Lawson proved that for any genus $g = 1, 2, \dots$, there is a compact minimal surface embedded in $\mathbb S^3$ \cite{lawson}. Examples of such surfaces were found in \cite{karcher} and \cite{kapouleas}. Lawson also conjectured that any compact toroidal minimal surface embedded in $\mathbb S^3$ is the Clifford torus \cite{lawson2}. This conjecture was proved in 2013 by Brendle \cite{brendle}.

In 2006 Butscher and Packard produced new examples of embedded, higher-genus CMC surfaces of $\mathbb S^3$ with small but non-zero mean curvature \cite{butscher}. A complete classification of CMC tori embedded in $\mathbb S^3$ was recently given by Andrews and Li \cite{AL}.

\section{Stability}
\label{stability}

The notion of stability of MOTS is related directly to the notion of being ``outermost''. In this paper we adhere to definitions introduced in \cite{ams, ams2}.

A marginally outer trapped surface $\mathcal S$ is called ``locally outermost'' in a given time slice $\Sigma$, if there exists a neighborhood $U$ of $\mathcal S$ such that the exterior part of $U$ does not contain any weakly outer trapped surface (a surface with nonpositive expansion $\theta_+$).

Andersson, Mars and Simon call a MOTS $\mathcal S$ ``stably outermost'', provided that there exists a function  $\psi \ge 0$, $\psi \neq 0$, on $\mathcal S$ such that  $\delta_{\psi m} \theta_+ \ge 0$. Here $\delta_{\psi m} \theta_+$ denotes the variation of $\theta_+$ with respect to the vector $\psi m^\mu$, and $m^\mu$ denotes a unit, outward-oriented vector normal to $\mathcal S$.

These definitions are directly related to the stability operator $L_\Sigma \psi = \delta_{\psi m} \theta_+$ of the form
\begin{eqnarray*}
L_\Sigma \psi & = & - \Delta_{\mathcal S} \psi + 2 s^A D_A \psi + \left( \frac{1}{2} R_{\mathcal S} - s_A s^A + D_A s^A \right. \\
& & \left. - \frac{1}{2} (\nabla_\mu l_+^\nu)(\nabla_\nu l_+^\mu) - G_{\mu \nu} l_+^\mu n^\nu \right) \psi.
\end{eqnarray*}
Following \cite{ams}, we use the symbols $D_A$ and $\Delta_{\mathcal S}$ for the covariant derivative and Laplacian with respect to the induced metric on $\mathcal S$, respectively. The coordinates on the surface $\mathcal S$ are denoted with capital Latin letters $A, B, \dots$ The vector $s_A$ is defined as
\[ s_A = -\frac{1}{2} g_{\alpha \beta} l_-^\beta \nabla_A l_+^\alpha, \]
and $G_{\mu\nu}$ is the Einstein tensor with respect to the 4-dimensional metric $g$.

It can be shown that the real parts of eigenvalues of the operator $L_\Sigma$ are bounded from below and that the principal eigenvalue (the eigenvalue with the smallest real part) of the operator $L_\Sigma$ is real. It can also be shown that the corresponding principal eigenfunction is either everywhere positive, or everywhere negative.

It was proved in \cite{ams} that $\mathcal S$ is stably outermost, if and only if the principal eigenvalue of $L_\Sigma$ is non-negative.

The fact that the toroidal MOTS described in this paper should be unstable is intuitive, and it is suggested by the SO(4) symmetry of the spatial part of the FLRW metric. Clearly, these MOTS are not ``locally outermost''. This follows directly from their construction. Consider the CMC Clifford tori, as described in the previous section. They form a one parameter family of CMC surfaces, parametrized by $\tau$. The value of $\theta_+$ is also constant at each of the tori, but it decreases in the outward direction, i.e., it is an increasing function of $\tau$. That means that a MOTS which belongs to this family is enclosed within other outer trapped surfaces.

Showing that these MOTS are unstable in the sense of \cite{ams} is more demanding, however computing the corresponding operator $L_\Sigma$ and looking at its properties is also an elegant and simple exercise, which demonstrates the power of the method introduced in \cite{ams} in a non-trivial setting.

Let us work in coordinates $(T,\sigma,\tau,\phi)$, as described in the previous section. The metric on the spacetime $\mathcal M$ is given by Eq.\ (\ref{gtoroidal}). Consequently, the induced metric on a $\tau = \mathrm{const}$ torus is simply
\[ h = S^2 ( \mathrm{sech}^2 \tau d\sigma^2 + \mathrm{tanh}^2 \tau d\phi^2), \]
and it is obviously flat. The corresponding scalar curvature $R_{\mathcal S}$ and connection coefficients vanish. In coordinates $(T,\sigma,\tau,\phi)$ the vectors $n^\mu$ and $m^\mu$ are
\[ n^\mu = (1, 0, 0, 0), \quad m^\mu = (0, 0,- \cosh \tau /S, 0). \]
A direct computation shows that $s_A = 0$. The remaining terms yield
\[ L_\Sigma \psi = - \frac{1}{S^2} \left[ (\cosh^2 \tau) \partial_{\sigma \sigma} \psi + (\coth^2 \tau) \partial_{\phi\phi} \psi \right] - B \psi, \]
where the constant $B$ is given by
\[ B = \frac{ \dot S^2 + (\dot S - \mathrm{csch} \tau)^2 + (\dot S + \sinh \tau)^2}{2 S^2} + \frac{3 (1 + \dot S^2)}{S^2}. \]

A standard separation of variables yields the spectrum of $L_\Sigma$ in the form
\[ \lambda = - B + \frac{1}{S^2} \left( l^2 \cosh^2 \tau + m^2 \coth^2 \tau \right), \quad l, m = 0, 1, 2\dots \]
Since $B$ is manifestly strictly positive, we conclude that the MOTS belonging to the family of CMC Clifford tori are unstable, as expected.

\section{Isoperimetric inequalities}

In \cite{khuri_xie} Khuri and one of us introduced some isoperimetric inequalities valid for toroidal surfaces. One of them, proved for toroidal minimal surfaces in the time-symmetric data, was already tested in our previous work \cite{kmmmx}. The family of CMC Clifford tori described in previous sections provides an opportunity to test such inequalities in a dynamical setting with $K_{ij} \neq 0$.

\begin{figure}
\includegraphics[width=\columnwidth]{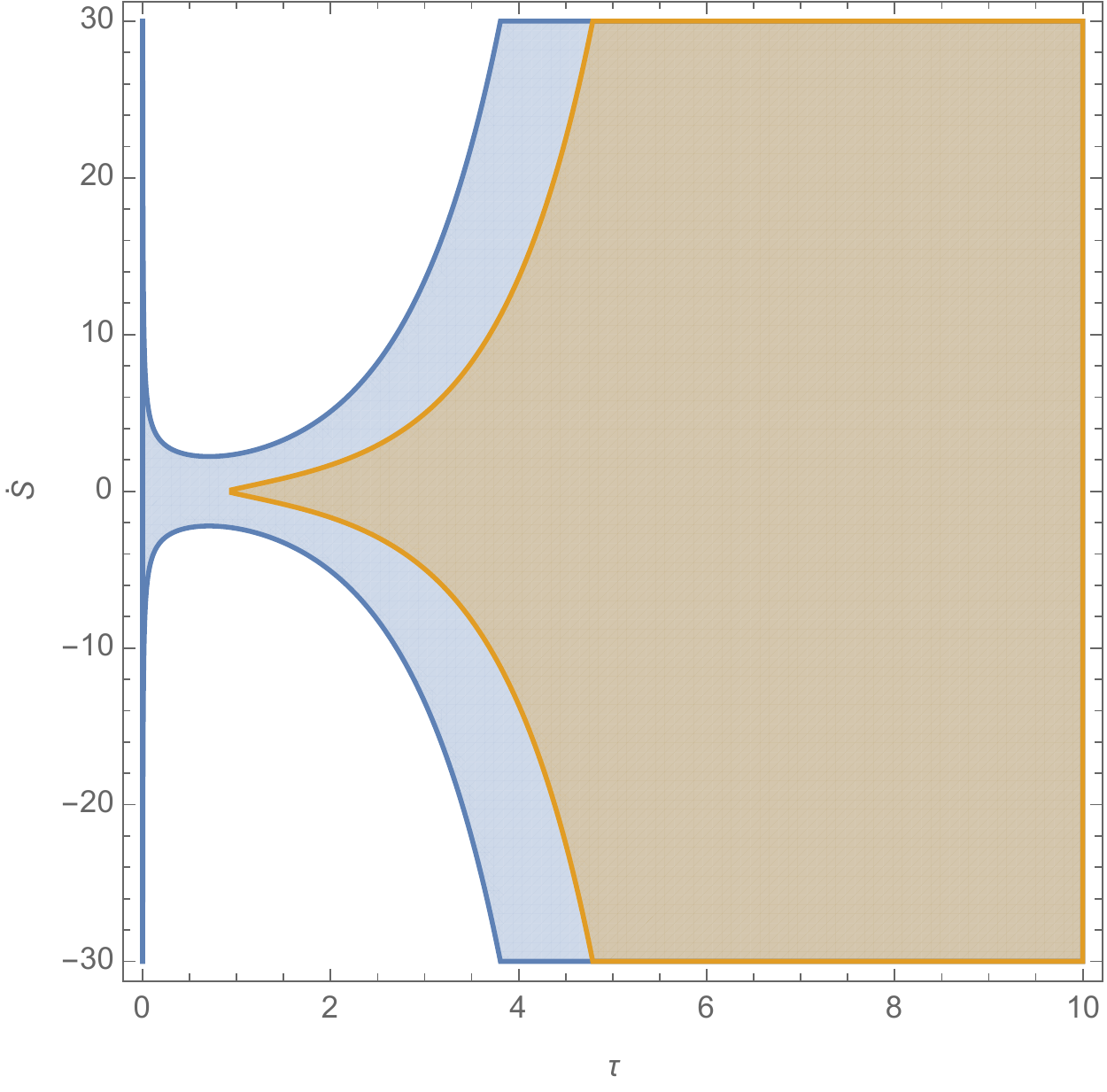}
\caption{\label{ineq_check}The larger region where (\ref{cond1}) is positive is marked in blue. The smaller region where (\ref{cond2}) is satisfied is marked in light brown.}
\end{figure}

The first inequality which we assess here applies to the time-symmetric case, corresponding to $\dot S = 0$ in our model. Let $\Omega$ denote the region inside a torus, and let $\partial \Omega$ denote its boundary. We define the rest-mass of the torus by $M = \int_\Omega dV \rho$, where the integration is taken with respect to the proper volume element. Define $L = \max_\Omega (\Phi^2 r)$, where $\Phi$ is the spatial conformal factor and $r$ denotes the cylindrical radius: $r = \sqrt{R^2 - z^2}$. This corresponds to the circumferential radius of the largest circle in the torus. Basing on \cite[][Eq.\ (5.3)]{khuri_xie} one expects for the ``untrapped'' tori with $H \ge 0$,
\begin{equation}
\label{ineq1}
\frac{2M}{\pi L} + \frac{1}{4 \pi^2 L} \int_{\partial \Omega} H ds \le 1.
\end{equation}
Let us choose $\Omega$ to be a region inside one of the CMC Clifford tori with a given parameter $\tau$. The volume of the torus and the corresponding rest-mass read $\int_\Omega dV = 2 \pi^2 S^3 \mathrm{sech}^2(\tau)$ and $M = \frac{3}{4} \pi S \mathrm{sech}^2 (\tau)$, respectively. The area of the surface of the torus is $\int_{\partial \Omega} ds = 4 \pi^2 S^2 \mathrm{sech}(\tau) \tanh (\tau)$, and the corresponding term $\int_{\partial \Omega} H ds = 2 \pi^2 S [\cosh(2 \tau) - 3] \mathrm{sech}^2(\tau)$. Finally $L = S$. For the left hand side of inequality (\ref{ineq1}) one obtains
\[ \frac{2M}{\pi L} + \frac{1}{4 \pi^2 L} \int_{\partial \Omega} H ds = \frac{1}{2} [1 + \tanh^2 (\tau)], \]
which is clearly smaller than one. It is quite surprising that inequality (\ref{ineq1}) turns out to be also valid for CMC tori with $H < 0$. For the minimal Clifford torus with $H = 0$ and $\tau = \tau_0$ one has
\[ \frac{2M}{\pi L} = \frac{3}{4}. \]
This result was already obtained in \cite{kmmmx}.

Let us now focus on the more general case with $K_{ij} \neq 0$. Suppose that the surface of the torus $\partial \Omega$ is ``strongly untrapped'', i.e., $H - |K_{ij}m^i m^j - \mathrm{tr} K| > 0$, where $m^i$ is the unit vector normal to $\partial \Omega$. In addition to $L$, let us also define $l = \min_\Omega (\Phi^2 r)$. According to \cite[][Eq.\ (2.16)]{khuri_xie} the following inequality is expected to hold
\begin{equation}
\label{ineq2}
M + \frac{L}{8 \pi l} \int_{\partial\Omega} (H - |K_{ij}m^i m^j - \mathrm{tr} K|) ds \leq \frac{\pi L^2}{2l}.
\end{equation}
For $\dot S \neq 0$ we have now $M = \frac{3}{4} \pi S (1 + \dot S^2) \mathrm{sech}^2 (\tau)$. An elementary computation yields $l = S \tanh(\tau)$. The second term in (\ref{ineq2}) reads
\begin{eqnarray*}
\lefteqn{\frac{L}{8 \pi l} \int_{\partial\Omega} (H - |K_{ij}m^i m^j - \mathrm{tr} K|) ds}\\
&& = \frac{1}{4} \pi  S \, \mathrm{csch}(\tau ) \mathrm{sech}(\tau ) [ -4 | \dot S |  \sinh (\tau ) + \cosh (2 \tau ) - 3].
\end{eqnarray*}
By collecting all terms together, one can now show that
\begin{eqnarray*}
\lefteqn{\frac{\pi L^2}{2l} - M - \frac{L}{8 \pi l} \int_{\partial\Omega} (H - |K_{ij}m^i m^j - \mathrm{tr} K|) ds} \\
&& = \frac{1}{4} \pi  S \, \text{sech}(\tau ) \left[ 4 (|\dot S| +\text{csch}(\tau ))-3 (\dot S^2 + 1 ) \text{sech}(\tau ) \right].
\end{eqnarray*}
Clearly, in order to confirm inequality (\ref{ineq2}) it suffices to inspect the sign of
\begin{equation}
\label{cond1}
 4 (|\dot S| +\text{csch}(\tau ))-3 (\dot S^2 + 1 ) \text{sech}(\tau ),
\end{equation}
which should be positive, provided that the surface of the torus with the given $\tau$ is strongly untrapped, that is for
\begin{equation}
\label{cond2}
\tau > \mathrm{arsinh} \left( |\dot S| + \sqrt{1 + \dot S^2} \right).
\end{equation}
The actual proof is elementary, but tedious. Figure \ref{ineq_check} shows that the region where the condition (\ref{cond2}) is satisfied is actually contained in the region where (\ref{cond1}) is positive. That means that inequality (\ref{ineq2}) is satisfied for a larger class of surfaces than the family containing only ``strongly untrapped'' ones.

\section{Concluding remarks}

Existence of spherical MOTS in FLRW geometries is well known, and there is a vast literature discussing different types of cosmological horizons (see \cite{faraoni} for a book review). Much less is known about non-spherically symmetric MOTS.

Since the 3-sphere $\mathbb S^3$ contains minimal and CMC surfaces of arbitrary genus, it is clear that one can have minimal surfaces and MOTS of any topology in closed FLRW spacetimes as well. This fact was already noticed in \cite{flores}. Constant mean curvature Clifford tori present in FLRW geometries have the advantage that they can be approached by a straightforward construction, allowing for a direct test of existing theorems concerning MOTS and toroidal surfaces in general relativity.

Similarly to spherical MOTS in FLRW spacetimes, the ones belonging to the class of CMC Clifford tori are also unstable. It is quite remarkable that the proof of this fact, which in general requires the knowledge of the sign of the principal eigenvalue of the complicated stability operator, can be accomplished by a simple, direct calculation. The fact that spherical MOTS in FLRW spacetimes are unstable in the sense of Andersson, Mars and Simon is shown in the Appendix.

\section*{Acknowledgments} We would like to thank Miko\l{}aj Korzy\'{n}ski, Edward Malec, and Walter Simon for discussions. P.\ Mach acknowledges the support of the Narodowe Centrum Nauki Grant No.\ DEC-2012/06/A/ST2/00397 and the hospitality of the School of Mathematical Sciences, Fudan University.  N.\ Xie is partially supported by the National Natural Science Foundation of China (Grants No.\ 11671089, No.\ 11421061).

\appendix*
\section{Instability of spherical MOTS in FLRW spacetimes}

In this Appendix we show that spherical MOTS in closed FLRW spacetimes are unstable in the sense of Andersson, Mars and Simon \cite{ams,ams2}. The calculation is essentially the same as the one presented in Sec.\ \ref{stability}, with a few minor changes. As before, it suffices to show that the principal eigenvalue of the corresponding stability operator is negative.

We work in spherical coordinates $(T,R,\theta,\varphi)$. The metric $g$ and the spatial conformal factor $\Phi$ are given by Eqs.\ (\ref{g1}) and (\ref{g2}), respectively. A 2-sphere $\mathcal S$ of constant radius $R$, embedded in a given time slice $\Sigma_t$, has a scalar curvature $R_\mathcal{S} = 2/(\Phi^4 R^2)$ and a mean curvature
\[ H = \frac{1}{\Phi^6 R^2} \partial_R (\Phi^4 R^2) = \frac{1 - C_1^2 R^2}{S C_1 R}. \]
The two null vectors $l_\pm^\mu$ read
\[ l_\pm^\mu = (1,\pm \Phi^{-2},0,0). \]
A direct computation shows that $s_A \equiv 0$. For the stability operator $L_\Sigma$ one obtains the expression
\[ L_\Sigma \psi = - \frac{1}{\Phi^4 R^2} \Delta_{\mathbb S^2} \psi - B \psi, \]
where $\Delta_{\mathbb S^2}$ denotes the Laplacian on the unit 2-sphere, and
\[ B = -\frac{1}{\Phi^4 R^2} + \frac{\dot S^2}{2 S^2} + \left( \frac{H}{2} + \frac{\dot S}{S} \right)^2 + \frac{3(1 + \dot S^2)}{S^2}. \]
This gives the spectrum of $L_\Sigma$ in the form
\[ \lambda = - B + \frac{l(l+1)}{\Phi^4 R^2}, \quad l = 0, 1, \dots \]
In order to establish the sign of $B$ we take into account that the sphere $\mathcal S$ is supposed to be a MOTS, i.e., $\theta_+ = H - 2 \dot S/S = 0$ on $\mathcal S$. This yields
\[ \lambda = - \frac{4 + 13 \dot S^2}{2 S^2} + \frac{1 + \dot S^2}{S^2} l (l+1), \quad l = 0, 1, \dots \]
The principal eigenvalue $\lambda = - (4 + 13 \dot S^2)/(2 S^2)$ is obviously negative.

\end{document}